\documentclass[aps,prl,twocolumn,superscriptaddress,showpacs]{revtex4}
\usepackage{graphicx}
\usepackage{amsmath}

\begin{document}

\title{Strong phonon-plasmon coupled modes in the graphene/silicon carbide heterosystem}

\author{R. J. Koch}
\affiliation{Institut f\"ur Physik and Institut fuer Mikro- und Nanotechnologien, TU Ilmenau, Germany}
\affiliation{Lehrstuhl f\"ur Technische Physik, Universit\"at Erlangen-N\"urnberg, Germany}

\author{Th. Seyller}
\affiliation{Lehrstuhl f\"ur Technische Physik, Universit\"at Erlangen-N\"urnberg, Germany}

\author{J. A. Schaefer}
\affiliation{Institut f\"ur Physik and Institut fuer Mikro- und Nanotechnologien, TU Ilmenau, Germany}
\affiliation{Department of Physics, Montana State University, USA}

\date{\today}

\begin{abstract}
We report on strong coupling of the charge carrier plasmon $\omega_{PL}$ in graphene with the surface optical phonon $\omega_{SO}$ of the underlying SiC(0001) substrate with low electron concentration ($n=1.2\times 10^{15}$ $cm^{-3}$) in the long wavelength limit ($q_\parallel \rightarrow 0$). Energy dependent energy-loss spectra give for the first time clear evidence of two coupled phonon-plasmon modes $\omega_\pm$ separated by a gap between $\omega_{SO}$ ($q_\parallel \rightarrow 0$) and $\omega_{TO}$ ($q_\parallel >> 0$), the transverse optical phonon mode, with a Fano-type shape, in particular for higher primary electron energies ($E_0 \ge 20eV$). A simplified model based on dielectric theory is able to simulate our energy - loss spectra as well as the dispersion of the two coupled phonon-plasmon modes $\omega_\pm$. In contrast, Liu and Willis \cite{willis2010} postulate in their recent publication no gap and a discontinuous dispersion curve with a one-peak structure from their energy-loss data.
\end{abstract}


\maketitle

The graphene silicon carbide heterosystem is a promising system for the future application of graphene in micro- and nanoelectronics \cite{berger2004,geim2004}. Silicon carbide as a substrate for microelectronics is already used industrially and the epitaxial growth of graphene on silicon carbide has already been investigated for several years now \cite{berger2004,angot2002}, and perfectionalized towards wafer scale homogeneous graphene \cite{emtsev2009}. Still, many of the interactions between the graphene and the silicon carbide substrate have yet to be understood. For example, the carrier dynamics may be strongly influenced by the long-range coupling to the polar modes of the substrate \cite{fratini2008}, which possibly results in a strong reduction of the graphene mobility, if compared to free standing graphene. This remote scattering can be important in future graphene devices.

In this contribution we report about our experimental investigation of the carriers in the conduction channel with the long-range polarization field created at the conductor/dielectric interface. Emphasis is also given to the theoretical interpretation of the experimental inelastic electron scattering results by calculating the dielectric surface loss function. The coupling of collective electron (or hole) modes with optical phonons in semiconductors (e.g. InN, InP, GaAs and others) has already been a target of extensive investigations and helped to understand important interface characteristics \cite{kloeckner2010,dubois1984,polyakov1996}. Unlike these conventional two dimensional electron gas systems (2DEG), graphene exhibits a linear electron dispersion relation, but the plasmon dispersion remains \cite{sarma2009}. Furthermore, the almost vanishing damping of the plasmon mode and the strong spatial confinement, in contrast to other sheet plasmons observed so far \cite{matz1981,petrie1994}, makes it a showcase model for the investigation of the coupled phonon plasmon modes.

\begin{figure}[b]
\includegraphics[width=6.5cm]{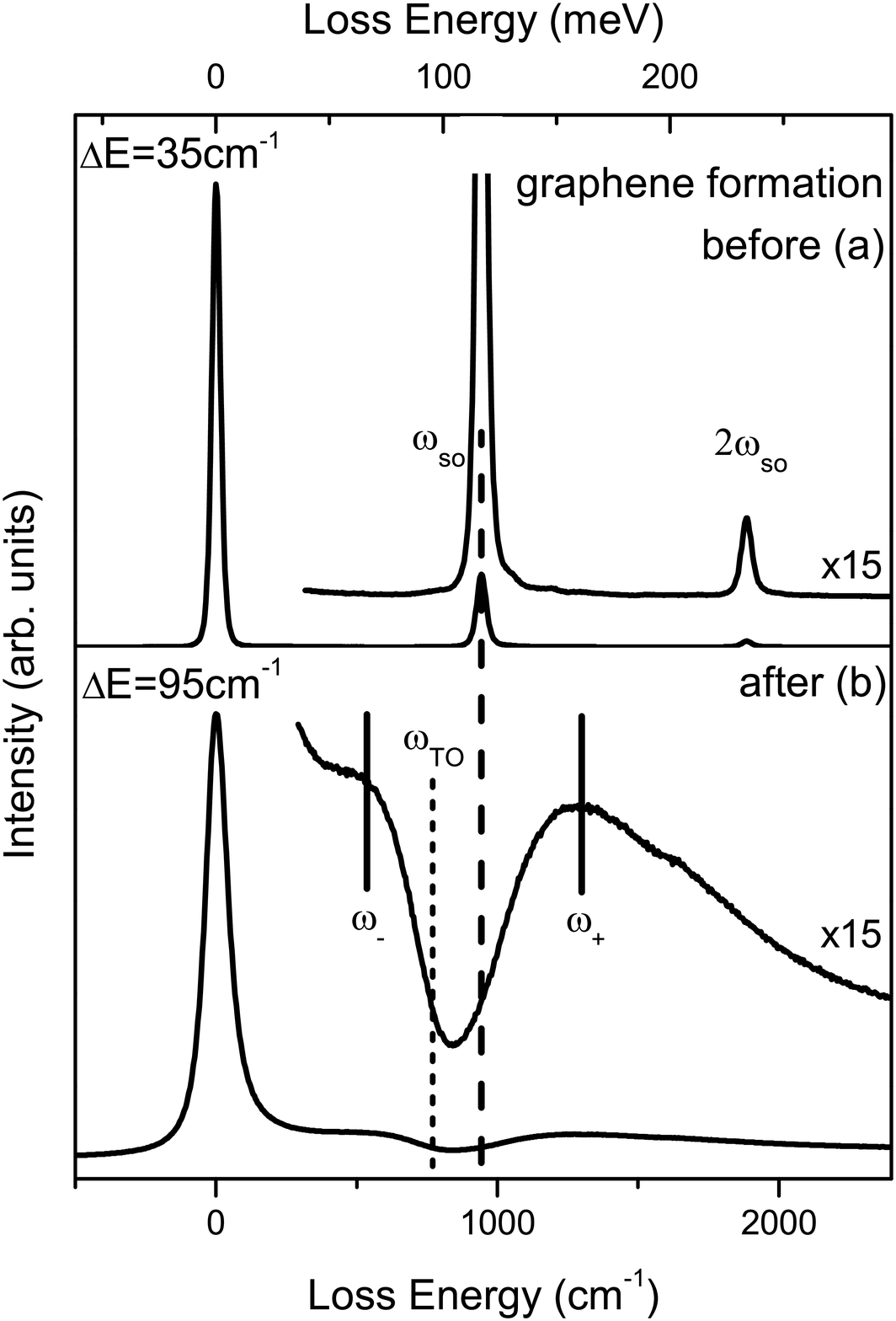}
\caption{\label{HREELSspecular} Energy-loss spectra in specular direction at a spectrometer resolution of $20$ $cm^{-1}$  on different samples. (a) on silicon carbide hydrogen etched with $E_0=36$ $eV$, (b) on the graphene/SiC(0001) heterosystem with $E_0 = 20$ $eV$. $T=300K$.
}
\end{figure}

Inelastic electron scattering utilizing special high resolution monochromators and analyzers, also known as high resolution electron energy loss spectroscopy (HREELS), is used in surface science to investigate intentional and unintentional adsorbates as well as surface phonon and plasmon modes on a wide variety of materials \cite{ibachmills}. Dispersion measurements can be obtained in different measurement methods. By changing the analyzer rotation angle impact scattered electrons can be analyzed over the whole Brillouin zone, which is mostly employed for measuring the phonon- and electron-dispersion\cite{langer2009}. To investigate the dispersion very close to the center of the Brillouin zone the analyzer is kept in specular direction and only the primary energy of the impinging electrons is varied. 

We took all HREELS spectra in specular direction, with the impact and scattering angle $\theta$ fixed at $64^\circ$ relative to the surface normal. The momentum transfer parallel to the surface for the dispersion measurements is calculated in this particular geometry from the impact energy E and the loss energy $\hbar \omega$ by 
\begin{equation}
\label{eqdisp}
 q_\parallel = \frac{\sqrt{2m_e}}{\hbar}sin(\theta) \left[ \sqrt{E} - \sqrt{E-\hbar\omega}\right]
\end{equation}
All experiments have been carried out in an UHV-system equipped with a HREELS spectrometer ``Delta 0.5'' originally designed by Ibach et al.\cite{ibachmills}. The pressure was kept below $1\times 10^{-10}$ Torr during all measurements.

The graphene/silicon carbide sample used for this investigation was prepared ex-\textit{situ} by a hydrogen etching step and the following graphitisation under atmospheric argon pressure \cite{emtsev2009}. This resulted in a 1.5 ML graphene in addition to the buffer layer measured by XPS (not shown here)\cite{emtsev2009}. After transfer to the ultrahigh vacuum (UHV) system the sample showed almost no contamination with hydrocarbons or water. In contrast, the hydrogen etched silicon carbide samples used for comparison, shows small amounts of hydrocarbons and dissociated and/or non dissociated water \cite{koch2010}.

\begin{figure}[t]
 \includegraphics[width=9cm,clip=true,viewport=1.5cm 0.5cm 30cm 21cm]{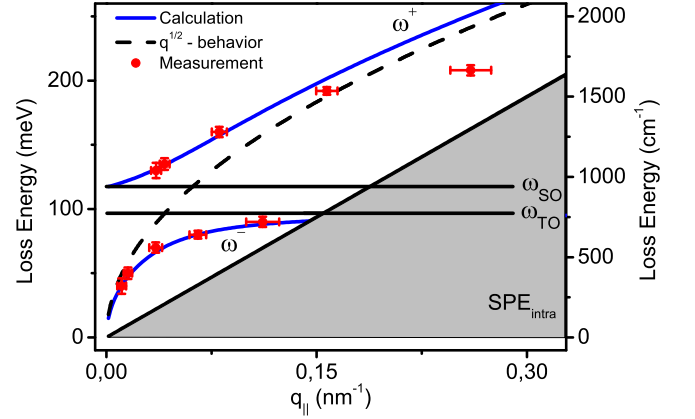}
 \caption{\label{HREELSdispersiongraph} Dispersion of the coupled phonon plasmon modes $\omega_\pm$. The red dots represent our measurements, taken at different primary beam energies ($2.5$ $eV\le E_0 \le 80$ $eV$), the blue lines result from theory (equation \ref{dispersionsformel}). The dashed line indicates a $\sqrt{q}$ behavior, which would be typical for a 2DEG. The gap between $\omega_{SO}$ and $\omega_{TO}$ is indicated by full horizonal lines (black). The shaded area represents single particle intraband excitations ($SPE_{intra}$).}
\end{figure}

Fig. \ref{HREELSspecular} represents an energy-loss spectrum in specular direction for 6H-SiC(0001) before (a) and after (b) graphene formation. As already shown \cite{koch2010} the surface optical phonons, also called Fuchs-Kliewer phonons \cite{fuchs1965}, are totally quenched. A two peak structure with maxima at $\omega_+=1270$ $cm^{-1}$ and $\omega_-=560$ $cm^{-1}$ results, which is shifted to lower (higher) energy values for higher (lower) primary beam energies and correspondingly smaller (higher) $q_\parallel$-values according to equation \ref{eqdisp} and our measured data, as shown in Fig. \ref{HREELSdispersiongraph} (red dots).

For the explanation of the dispersion shown in Fig. \ref{HREELSdispersiongraph} (blue line) we consider a thin conducting layer with a charge carrier plasmon, brought on top of a substrate with strong surface optical phonons with $\omega_{SO}=\omega_{TO} \sqrt{\frac{1+\epsilon_0}{1+\epsilon_\infty}}=945$ $cm^{-1}$. These two modes couple, which can be described in the framework of dielectric theory \cite{koch2010,dubois1984,balster1998,polyakov1996}. The dielectric function of a phonon is given by 
\begin{equation}
 \epsilon (\omega)=\epsilon_\infty + \frac{(\epsilon_0-\epsilon_\infty)\omega_{TO}^2}{\omega_{TO}^2-\omega^2-i\gamma\omega}
 \label{phonon}
\end{equation}
and the dielectric function of a volume charge carrier plasmon by
\begin{equation}
 \epsilon (\omega)= \epsilon_\infty - \frac{\omega_{PL}^2}{\omega^2+i\Gamma\omega}
 \label{plasmon}
\end{equation} where the plasmon frequency is a constant, which is determined by the charge carrier density and material properties as discussed below.\\
The description as a volume plasmon is not sufficent here, a two dimensional plasmon mode depending on the thickness d and showing a dispersion of $\omega_{P2D} (q_\parallel) \propto \sqrt{q_\parallel}$ has to be taken into account:
\begin{equation}
 \omega_{P2D}=\omega_{PL} \sqrt{\frac{q_\parallel d}{1+\epsilon_\infty^{SiC}}}
 \label{plasmondispersion}
\end{equation}
Neglegting phonon and plasmon damping ($\gamma = 0$ and $\Gamma = 0$) and restricting the calculation to the region where $q_\parallel d <<1$ allows the determination of an exact expression for the dispersion of the coupled modes \cite{dubois1984}:
\begin{multline}
 \omega^2_\pm (q_\parallel)=\frac{1}{2}\left[ \omega_{SO}^2 + \omega_{P2D}^2 (q_\parallel) \right]
				      \pm \frac{1}{2} \Big\lbrace \left[ \omega_{SO}^2 - \omega_{P2D}^2 (q_\parallel) \right]^2 \\+ 4 \left[ \omega_{SO}^2 - \omega^2_{TO} \right] \omega_{P2D}^2 (q_\parallel) \Big\rbrace^\frac{1}{2}
\label{dispersionsformel}
\end{multline}
These coupled modes have mixed phonon and plasmon characteristics. Referring to Fig. \ref{HREELSdispersiongraph}, at low momentum transfer ($q_\parallel\rightarrow 0$) the $\omega_+$ mode behaves more like a phonon mode and converges to the frequency of the classical surface optical phonon $\omega_{SO}$. The $\omega_-$ mode vanishes to zero in the long wavelength limit and behaves similar to a classical two dimensional plasmon mode. On the other hand, at high momentum transfers the $\omega_-$ mode converges to the frequency of the TO phonon and the $\omega_+$ mode behaves like a plasmon. It has to be noted, that this anticrossing also exhibits an energetically forbidden zone between the TO phonon frequency and the surface optical phonon frequency. 

Fig. \ref{HREELSdispersiongraph} displays the resulting dispersion together with the theoretical dispersion of the two coupled modes, which follow from equation \ref{dispersionsformel}. A plasmon frequency of $12000$ $cm^{-1}$ was fitted and a TO phonon frequency of $760$ $cm^{-1}$ was used \cite{harris1995}. Except for the $\omega_+$ mode at high momentum transfers the curves fit perfectly to the measured data. The reason for the discrepancy at high momentum transfers lies in the non sufficient satisfaction of the condition $q_\parallel d <<1$.

In contrast to material systems investigated earlier (e.g. silver on GaAs) \cite{dubois1984}, in the graphene on silicon carbide heterosystem this dispersion relation can be directly verified (see Fig. \ref{HREELSdispersiongraph}). In other systems so far, the plasmon damping is always notably higher than in the graphene silicon carbide system. Therefore the dispersion relation calculated did not fit the dispersion observed in these systems. In particular, here no $\omega_{SO}$ mode is visible, which arises as a third solution in eq. \ref{dispersionsformel}, when taking finite damping into account. A closer look on the plasmon damping can be gained by simulation of the whole HREELS spectrum.

\begin{figure}[t]
 \includegraphics[width=8cm]{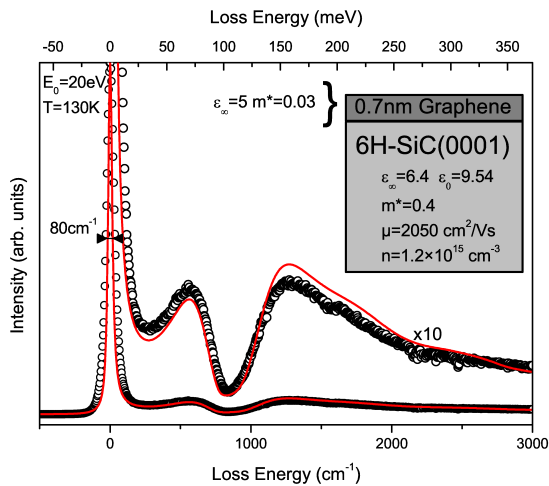}
 \caption{\label{HREELSsimulation} HREELS spectrum (open circles) taken at 20eV primary beam energy together with simulation (red line). The inset gives details of the two layer model. $T=130K$.}
\end{figure}

It is well known from dipole scattering theory\cite{ibachmills,schaich1981}, that the energy loss probability in a HREELS experiment is given by 
\begin{equation}
 P(\omega)=\int\limits_{\{ q_\parallel \}} \frac{q_\parallel d^2 q_\parallel}{\left[ v^2_\perp q_\parallel^2 + (\omega-v_\parallel q_\parallel)^2 \right]} \Im\left\lbrace \frac{-1}{\epsilon(\omega)+1} \right\rbrace
\end{equation}
The integration limits of $q_\parallel$ are determined by the angular dimension of the spectrometer aperture. Convoluting this loss probability with a suitable spectrometer function, also taking into account temperature effects and double losses allows us to directly compare the simulations with the measured spectra\cite{lambin1990}.

The effective dielectric function to describe the graphene silicon carbide heterosystem is build out of the elementary dielectric functions of the phonon and plasmon contributions (see eqs. \ref{phonon} and \ref{plasmon}). The model includes two layers, one for the silicon carbide substrate and one for the graphene overlayer. The dielectric function for the graphene overlayer just includes the charge carrier plasmon and is therefore identical to eq. \ref{plasmon}. For the silicon carbide layer both, a charge carrier plasmon from the slightly n-doped silicon carbide beneath ($n\approx 1.2 \times 10^{15} cm^{-3}$) and the TO phonon, have been included:
\begin{equation}
 \epsilon_{SiC}(\omega)=\epsilon_\infty + \frac{(\epsilon_0-\epsilon_\infty)\omega_{TO}^2}{\omega_{TO}^2-\omega-i\gamma\omega} - \frac{\omega_{PL}^2}{\omega^2+i\Gamma_{SiC}\omega}
\end{equation}
The effective dielectric function, which in contrast to the single layer dielectric functions now also takes the parallel momentum transfer $q_\parallel$ into account, can now be calculated \cite{balster2006,lambin1985}:
\begin{multline}
 \epsilon (\omega,q_\parallel)=\epsilon_{graphene}(\omega) coth(q_\parallel d)-\\ \frac{\left[ \dfrac{\epsilon_{graphene}(\omega)}{sinh(q_\parallel d)} \right]^2}{\epsilon_{graphene}(\omega) coth(q_\parallel d) + \epsilon_{SiC}(\omega)}
\end{multline}
\begin{table}[b]
\caption{\label{ergebnisse} Parameters for the simulation of the graphene silicon carbide heterostructure HREELS spectrum}
 \begin{tabular}{l|ccccc}
  &d&$\omega_{TO}$ $(cm^{-1})$&$\gamma_{TO}$&$\omega_{PL}$ $(cm^{-1})$&$\Gamma_{PL}$\\ \hline
  Graphene&0.7 nm&&&$12000$ &$1.5\%$ \\
  SiC&$\infty$&$760$&$0.4\%$&$16$&$70\%$
  
 \end{tabular} 
\end{table} 
Although the basic theory behind this calculation is similar to the simple dispersion calculation shown above, it allows us to take the plasmon and phonon damping into account. 

Fig. \ref{HREELSsimulation} shows the numerical simulation (red line) together with the corresponding HREELS-spectrum (open circles). The simulation fits the measured data quite well, even though no band bending or interface effects (i.e. change of the charge carrier concentration in the silicon carbide as a function of depth) have been taken into account. For other primary beam energies apart from the $E_0=20eV$-spectrum shown in Fig. \ref{HREELSsimulation} simulations were carried out, too (not shown here). These simulations do fit the experimental results also very well. In general, the higher the primary beam energy is chosen, the better the simulation fits the experimental data. In addition it should be mentioned that with higher primary beam energies the $\omega_+$ mode exhibits a Fano type resonance shape \cite{fano1961}. Further discussion of this behavior is beyond the scope of this contribution. 

Table \ref{ergebnisse} summarizes the parameters used in the simulation. The plasmon frequency is about $500$ $cm^{-1}$ lower than in the dispersion simulation. This can be attributed to taking the damping into account.

From angle resolved photoelectron spectroscopy it is well known that graphene in the graphene silicon carbide heterosystem is electron doped. The measurements \cite{bostwick2007,bostwick2008, starke2009} show, that the linearly dispersing $\pi$-bands which provide the free charge carriers for the plasmons observed start about $E_F=450 meV$ below the Fermi level of the heterosystem. Using the relations $E_F=\hbar v_F k_F$ and $k_F=\sqrt{4\pi n/(g_sg_v)}$ (with $g_s=2$ and $g_v=2$ for the spin and valley degeneracies)\cite{neto2009}, the sheet electron density is determined to be $1.5\times 10^{13} cm^{-2}$. If we consider a typical plasmon which is found in a two dimensional electron gas by setting $\omega_{PL}=\sqrt{\frac{n e^2}{\epsilon m^*}}$, implying a parabolic component in the band structure, we obtain an effective mass of $0.03\times m_e$. The plasmon damping $\Gamma_{PL}=e/(\mu m^*)$ can be used to calculate the electron mobility $\mu=1700 cm^2/(Vs)$. These values fit the picture of previously published values from Hall mobility measurements \cite{emtsev2009}. 

On the other hand, Das Sarma et al. \cite{sarma2009} calculated the plasmon frequency for graphene depending on the electron density for a two dimensional dirac system to be:
\begin{equation}
 \omega_{P2D}=\sqrt{\frac{e^2}{\epsilon_0 \hbar v_F}} \left( 4\pi\cdot n_{2D} \right)^{1/4} v_F \sqrt{q_\parallel}
\end{equation}
This model for the plasmon frequency takes the linear band dispersion in graphene into account. Using the experimental data fit derived with eq. \ref{plasmondispersion} one can calculate the electron density to be $1.4 \times 10^{11} cm^{-2}$. For this density the Fermi energy shift would only be $E_F=45 meV$, one order of magnitude lower than measured. The doping of graphene on silicon carbide seems to be too high, so that a nonlinear contribution to the electronic dispersion has already to be considered.

We have shown, that the strong coupling between the charge carrier plasmon in graphene and the surface optical phonon in silicon carbide can be understood in terms of dielectric theory. The dispersion differs strongly from the phonon and plasmon dispersion in an uncoupled case. This understanding of the coupled modes is especially important for transport calculations in the epitaxial graphene silicon carbide heterosystem. The scattering processes with the coupled plasmon phonon modes, in particular in the low momentum low energy regime strongly depend on the changed dispersion of the $\omega_-$ branch of the dispersion.

\begin{acknowledgments}
Technical assistance by T. Haensel and G. Hartung, fruitful discussions with S. das Sarma and H. Hwang as well as the continuous support by the German Science Foundation(DFG) are greatfully acknowledged.
\end{acknowledgments}

\bibliography{bibliographie}

\end{document}